\documentclass[12pt, a4paper]{article}
\usepackage[latin1]{inputenc}
\usepackage[dvips]{graphicx,color}
\usepackage{amsmath}
\usepackage{amsfonts}
\usepackage{amssymb}
\begin{document}
% Put preprint number in top-right.
\def\pplogo{\vbox{\kern-\headheight\kern -29pt
\halign{##&##\hfil\cr&{\ppnumber}\cr\rule{0pt}{2.5ex}&\ppdate\cr}}}
\makeatletter
\def\ps@firstpage{\ps@empty \def\@oddhead{\hss\pplogo}%
  \let\@evenhead\@oddhead % in case an article starts on a left-hand page
}%      The only change in \maketitle is \thispagestyle{firstpage} instead of \thispagestyle{plain}
\def\maketitle{\par
 \begingroup
 \def\thefootnote{\fnsymbol{footnote}}
 \def\@makefnmark{\hbox{$^{\@thefnmark}$\hss}}
 \if@twocolumn
 \twocolumn[\@maketitle]
 \else \newpage
 \global\@topnum\z@ \@maketitle \fi\thispagestyle{firstpage}\@thanks
 \endgroup
 \setcounter{footnote}{0}
 \let\maketitle\relax
 \let\@maketitle\relax
 \gdef\@thanks{}\gdef\@author{}\gdef\@title{}\let\thanks\relax}
\makeatother

%%%%%%%%%%%%%%%%%%%%%%%%%%%%%%%%%%%%%%%%%%%%%%%%

\setcounter{page}0
\def\ppnumber{\vbox{\baselineskip14pt
%\hbox{hep-th/0000000}
}}
\def\ppdate{RUNHETC-2006-25} \date{}

\author{\\Gonzalo Torroba\footnote{torrobag@physics.rutgers.edu}\\
[7mm]
{\normalsize Department of Physics and Astronomy, Rutgers University}\\
{\normalsize Piscataway, NJ 08855, U.S.A.}\\}

\title{\bf \LARGE {Finiteness of Flux Vacua from \\Geometric Transitions}}
\maketitle
\vskip 2cm
\begin{abstract} \normalsize
\noindent We argue for finiteness of flux vacua around type IIB CY
singularities by computing their gauge theory duals. This leads us
to propose a geometric transition where the compact 3-cycles
support both RR and NS flux, while the open string side contains
5-brane bound states. By a suitable combination of S duality and
symplectic transformations, both sides are shown to have the same
IR physics. The finiteness then follows from a holomorphic change
of couplings in the gauge side. As a nontrivial test, we compute
the number of vacua on both sides for the conifold and the
Argyres-Douglas point, and we find perfect agreement.
\end{abstract}
\bigskip
\newpage

\tableofcontents

\vskip 1cm
%%%%%%%%%%%%%%%%%%%%%%%%%%%%%%%%%%%%%%%%%%%%%%%%%%%%%%%%%%%%%%%%%%%%%
%%%%%%%%%%%%%%%%%%%%%%%%%%%%%%%%%%%%%%%%%%%%%%%%%%%%%%%%%%%%%%%%%%%%%
%%%%%%%%%%%%%%%%%%%%%%%%%%%%%%%%%%%%%%%%%%%%%%%%%%%%%%%%%%%%%%%%%%%%%
%%%%%%%%%%%%%%%%%%%%%%%%%%%%%%%%%%%%%%%%%%%%%%%%%%%%%%%%%%%%%%%%%%%%%
%%%%%%%%%%%%%%%%%%%%%%%%%%%%%%%%%%%%%%%%%%%%%%%%%%%%%%%%%%%%%%%%%%%%%
\section{Introduction}\label{sec:intro}

The study of string vacua in flux compactifications of type IIB
has attracted much attention, in part as a setting where many
properties of the landscape of vacua are under control
(see~\cite{review} for a recent review). One of the first issues
to be addressed here is whether the number of realistic string
vacua is finite~\cite{finite}. Recently it has been shown in
~\cite{lu} that the Ashok-Douglas index of supersymmetric vacua is
finite. This is a crucial step towards proving the finiteness of
flux vacua.

The aim of the present work is to understand the physics
underlying the previous result. It was shown in~\cite{ashok} that
the index of supersymmetric vacua is finite around smooth points
in moduli space; the analysis may be restricted then to
singularities of the moduli space where the curvature diverges.
The finiteness proof~\cite{lu} is based on Weil-Petersson geometry
and a detailed analysis of degenerations of Hodge structures on
the moduli space. However, from the string theory point of view it
is not clear which is the physical mechanism responsible for this.
Specially, why singularities leading to very different field
theories all give a finite number of vacua.

Our approach may be summarized as follows. We construct
Calabi-Yau's where singularities are easily embedded and argue for
finiteness of vacua around them by computing their dual gauge
theories. We establish a precise correspondence between flux and
gauge degrees of freedom. This shows that the gauge theories are
generalized versions of the ones obtained through the
Dijkgraaf-Vafa correspondence~\cite{dv}; but they still have
finitely many vacua. As we shall see, the underlying reason for
this is the topological nature of the chiral ring of such
theories.

In section \ref{sec:generalcy} we discuss the
type IIB noncompact model that can embed ADE singularities and
study the nonperturbative superpotential generated by fluxes.
Next, in section \ref{sec:fluxcurve} we derive the formula for
counting vacua in the previous setup; this involves nontrivial
steps because of the noncompact nature of the model. Then in section
\ref{sec:dualym} we construct the dual gauge theory after the geometric
transition, applying S-duality to the Dirac-Born-Infeld (DBI) action. The
field theory turns
out to be a generalization of the usual $N=1$ SYM encountered in
geometric transitions. The argument for finiteness of vacua is
presented in section \ref{sec:finite}; it is based on the
holomorphic dependence of the gauge effective superpotential on
nondynamical fields (couplings). Finally, in section
\ref{sec:examples} we show that the computations from the gravity
and gauge side agree for the conifold and Argyres-Douglas
singularities. Section \ref{sec:concl} contains our conclusions.

%%%%%%%%%%%%%%%%%%%%%%%%%%%%%%%%%%%%%%%%%%%%%%%%%%%%%%%%%%%%%%%%%%%%%
%%%%%%%%%%%%%%%%%%%%%%%%%%%%%%%%%%%%%%%%%%%%%%%%%%%%%%%%%%%%%%%%%%%%%
%%%%%%%%%%%%%%%%%%%%%%%%%%%%%%%%%%%%%%%%%%%%%%%%%%%%%%%%%%%%%%%%%%%%%
%%%%%%%%%%%%%%%%%%%%%%%%%%%%%%%%%%%%%%%%%%%%%%%%%%%%%%%%%%%%%%%%%%%%%
%%%%%%%%%%%%%%%%%%%%%%%%%%%%%%%%%%%%%%%%%%%%%%%%%%%%%%%%%%%%%%%%%%%%%
\section{Fluxes in noncompact Calabi-Yau's}\label{sec:generalcy}

We start by studying moduli stabilization in type IIB theory in a
Calabi-Yau threefold. Since we are interested in analyzing a
neighborhood of an ADE singularity, it is enough to consider
noncompact threefolds of the form
\begin{equation}\label{eq:cy1}
P:= u^2+v^2+F(x,y)=0\,;
\end{equation}
the nontrivial dynamics comes from the complex curve $\Sigma$:
$F(x,y)=0$. (\ref{eq:cy1}) may be thought as a decoupling limit
$M_{Pl} \to \infty$ of an adequate compact variety~\cite{self},
although this will not be necessary for our purposes.

For concreteness, let us consider the case of a hyperelliptic
curve where we can realize singularities of the A-type:
\begin{equation}\label{eq:cy2}
F(x,y)=y^2-W'(x)^2-f_{n-1}(x)=0\,.
\end{equation}
$W'(x)$ is a polynomial of degree $n$, and will play the role of
the superpotential in the gauge theory:
\begin{equation}\label{eq:W1}
W'(x)=g_n \prod_{i=1} ^n (x-a_i)\,.
\end{equation}
$f_{n-1}(x)=\sum_{k=1}^n f_k x^{k-1}$ is a deformation of the
singular curve $y^2=W'(x)^2$. Its effect is to split $a_i \to
(a_i^-\,,a_i^+)$. If all the roots of $W$ are different then the
singular curve has just ODP (conifold) singularities. We will also
encounter more complicated singularities, where three or more
roots coincide.

The fact that (\ref{eq:cy2}) is the same variety
that appears in the Dijkgraaf-Vafa duality~\cite{dv} is not a
coincidence; the (generalized) gauge dual will play a major role
in proving the finiteness of the number of vacua. Furthermore, such local string models
have been considered recently in the context of soft supersymmetry breaking~\cite{localsusy}.

For our future computations, it is crucial to remark the
following. In the four dimensional effective field theory (EFT), the moduli $(a_i, f_k)$
have a very different interpretation. Fluctuations in $a_i$ have
infinite energy and hence are non-dynamical; each arbitrary choice
of $a_i$ will give a different 4d theory so they can be
interpreted as couplings. On the other hand, the $f_k$'s are
dynamical and are interpreted as scalar fields in vector
multiplets. Their gauge theory meaning will become clear in
section \ref{sec:dualym}.

As shown in \cite{intri}, the periods of the noncompact threefold
reduce to periods of the hyperelliptic curve:
\begin{equation}\label{eq:periods2}
S_i=\int_{A_i} R(x)dx\;\;,\;\;\frac{\partial \mathcal F}{\partial
S_i}=2\pi i \int_{B_i} R(x) dx\,
\end{equation}
with
\begin{equation}\label{eq:R}
2R(x)= W'(x)-\sqrt{W'(x)^2+f_{n-1}(x)}\,.
\end{equation}
The cycle $A_i$ surrounds the cut $[a_i^-, \,a_i^+]$; $B_i$ is the
noncompact cycle dual to $A_i$, running from $x=a_i$ to infinity.
The $B$-periods need to be regulated; this will be discussed
shortly. Therefore all the computations can be done directly on
the hyperelliptic curve $y(x)$ of genus $g=n-1$.

When $x \to \infty$
\begin{equation}\label{eq:Rtoinf}
R(x) \to -\frac{f_n}{2g_n x}\,.
\end{equation}
This implies that $R$ (and $y$) is a differential of the third
kind on $\Sigma$~\cite{farkas}. For any value $x \in \mathbb C$,
there are two points on the Riemann surface $\Sigma$; let $P,
\tilde P \in \Sigma$ denote the points corresponding to $x =
\infty$. Then $R(x) dx$ is a holomorphic differential only on the
punctured surface $\Sigma'=\Sigma-\{P, \tilde P\}$.

\begin{figure}[tbf]
\centering
\begin{picture}(0,0)%
\includegraphics[scale=0.5]{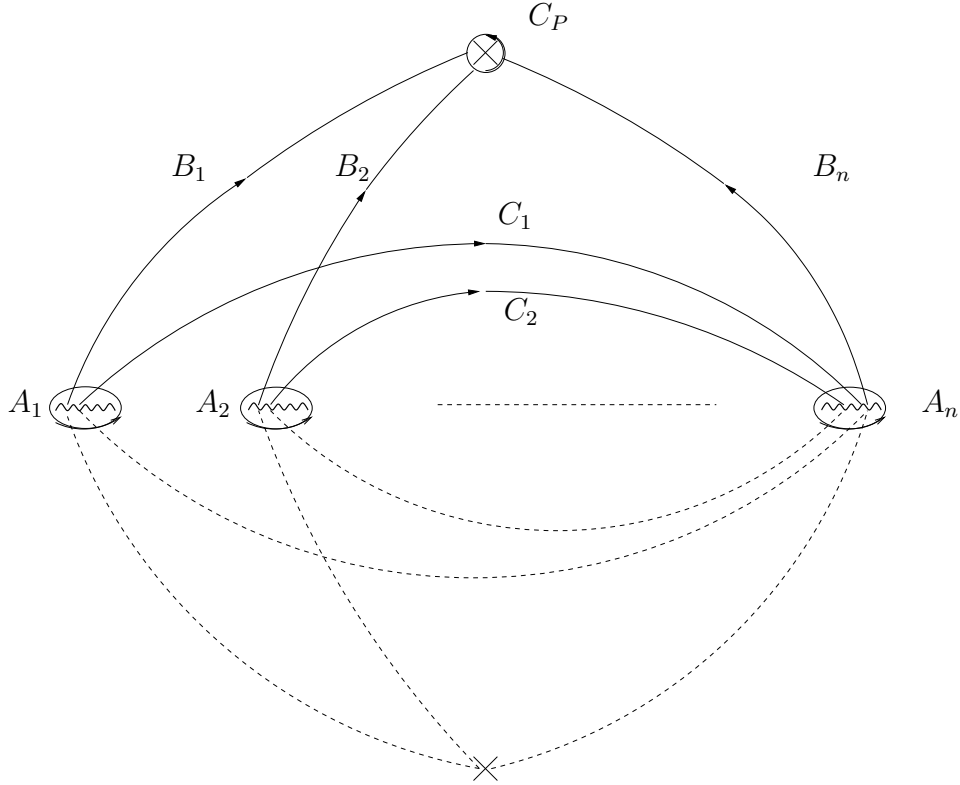}%
\end{picture}%
\setlength{\unitlength}{2000sp}%3947*scale
\begingroup\makeatletter\ifx\SetFigFont\undefined%
\gdef\SetFigFont#1#2#3#4#5{%
  \reset@font\fontsize{#1}{#2pt}%
  \fontfamily{#3}\fontseries{#4}\fontshape{#5}%
  \selectfont}%
\fi\endgroup%
\begin{picture}(11325,9543)(601,-8923)
\put(601,-4336){\makebox(0,0)[lb]{\smash{\SetFigFont{12}{14.4}{\rmdefault}{\mddefault}{\updefault}{\color[rgb]{0,0,0}$A_1$}%
}}}
\put(2926,-4336){\makebox(0,0)[lb]{\smash{\SetFigFont{12}{14.4}{\rmdefault}{\mddefault}{\updefault}{\color[rgb]{0,0,0}$A_2$}%
}}}
\put(11926,-4336){\makebox(0,0)[lb]{\smash{\SetFigFont{12}{14.4}{\rmdefault}{\mddefault}{\updefault}{\color[rgb]{0,0,0}$A_n$}%
}}}
\put(7051,464){\makebox(0,0)[lb]{\smash{\SetFigFont{12}{14.4}{\rmdefault}{\mddefault}{\updefault}{\color[rgb]{0,0,0}$C_P$}%
}}}
\put(2626,-1411){\makebox(0,0)[lb]{\smash{\SetFigFont{12}{14.4}{\rmdefault}{\mddefault}{\updefault}{\color[rgb]{0,0,0}$B_1$}%
}}}
\put(4651,-1411){\makebox(0,0)[lb]{\smash{\SetFigFont{12}{14.4}{\rmdefault}{\mddefault}{\updefault}{\color[rgb]{0,0,0}$B_2$}%
}}}
\put(10576,-1411){\makebox(0,0)[lb]{\smash{\SetFigFont{12}{14.4}{\rmdefault}{\mddefault}{\updefault}{\color[rgb]{0,0,0}$B_n$}%
}}}
\put(6676,-2011){\makebox(0,0)[lb]{\smash{\SetFigFont{12}{14.4}{\rmdefault}{\mddefault}{\updefault}{\color[rgb]{0,0,0}$C_1$}%
}}}
\put(6751,-3200){\makebox(0,0)[lb]{\smash{\SetFigFont{12}{14.4}{\rmdefault}{\mddefault}{\updefault}{\color[rgb]{0,0,0}$C_2$}%
}}}
\end{picture}
\caption{\small Homology elements of $\Sigma$ and $\Sigma'$.}
\label{fig:homology}
\end{figure}

The details of the homology of $\Sigma$ and the effect of the
punctures were considered in~\cite{csw1} and we follow their
conventions. A choice of homology cycles is shown in Figure
\ref{fig:homology}; $B_j$ runs through the $j$-th cut, from
$\tilde P$ to $P$. From these noncompact cycles we construct
$C_i=B_i-B_n$. Besides, $C_P$ and $C_{\tilde P}$ circle the
punctures at $P$ and $\tilde P$ respectively. The canonical
symplectic basis of $\Sigma$ is $(A_i, C_j)$, $i,j=1,\ldots,
g=n-1$. In $\Sigma$, $A_1+\ldots+A_n \equiv 0$ so $A_n$ is not
independent; however, in $\Sigma'$, $A_1+\ldots+A_n =-C_P$. This
means that we can take $A_n$ to be an independent cycle and use
this to fix the values of the meromorphic differentials at
infinity. A symplectic basis for $H_1(\Sigma', \mathbb Z)$ is
hence $(A_i, B_j)$, $i,j=1,\ldots,n$.

In the holomorphic decomposition $H^1(\Sigma, {\mathbb
C})=H^{1,0}(\Sigma, {\mathbb C})+H^{0,1}(\Sigma, {\mathbb C})$,
there is a unique basis of holomorphic differentials
~\cite{farkas} $(\zeta_1,\ldots,\zeta_g)$ such that
\begin{equation}\label{eq:zeta}
\int_{A_j} \zeta_k=\delta_{jk}\;\;,\;\; {\rm Im}\, \Pi \ge 0
\end{equation}
where the period matrix $\Pi$ is defined to be the symmetric
matrix
$$
\Pi_{jk}=\int_{C_j} \zeta_k\,.
$$
They can be constructed as linear combinations of the
differentials
\begin{equation}\label{eq:diff1}
\frac{\partial}{\partial f_k}\,y
\,dx=\frac{x^{k-1}}{2y}\,dx\;\;,\;\;k=1,\ldots,n-1\,.
\end{equation}
The third kind differential
\begin{equation}\label{eq:diff2}
g_n \frac{\partial}{\partial f_n}\,y \,dx=\frac{g_n
x^{n-1}}{2y}\,dx
\end{equation}
has residues $\pm 1$ at $P, \tilde P$ respectively. An adequate
linear combination of (\ref{eq:diff1}) and (\ref{eq:diff2}) will
give the unique third kind differential $\tau_{P,\, \tilde P}$
such that
$$
{\rm ord}_P \,\tau_{P, \,\tilde P}= {\rm ord}_{\tilde P}
\,\tau_{P, \,\tilde P}=-1\,,
$$
$$
{\rm res}_P \,\tau_{P, \,\tilde P}= 1\;,\;\; {\rm res}_{\tilde P}
\,\tau_{P, \,\tilde P}= -1\,.
$$
Every holomorphic differential on $\Sigma'$ can be written as a
linear combination of $(\zeta_1,\ldots,\zeta_g,\tau_{P \tilde
P})$. Such differentials are meromorphic differentials on $\Sigma$
with at most simple poles. A more symmetric description follows
from taking $A_n$ (instead of $C_P$) to be an independent cycle;
hence the basis of allowed differentials will be
$(\zeta_1,\ldots,\zeta_{n-1},\,\zeta_n)$ where $\zeta_n$ is a
superposition of $\zeta_i$ and $\tau_{P,\,\tilde P}$ fixed by
$\int_{A_j} \,\zeta_n=\delta_{jn}$, $j=1,\ldots,\,n$.

%%%%%%%%%%%%%%%%%%%%%%%%%%%%%%%%%%%%%%%%%%%%%%%%%%%%%%%%%%%%%%%%%%%%%
%%%%%%%%%%%%%%%%%%%%%%%%%%%%%%%%%%%%%%%%%%%%%%%%%%%%%%%%%%%%%%%%%%%%%
\subsection{Superpotential and fluxes}\label{subsec:superp}

The complex moduli of $X$ are stabilized by turning on 3-form
fluxes $G_3 := F_3 - \tau H_3$, which generate the nonperturbative
superpotential~\cite{gvw}
\begin{equation}\label{eq:superp1}
W_{eff}=\int_X\,G_3 \wedge \Omega\,.
\end{equation}
In the noncompact model, the axio-dilaton $\tau$ is fixed,
corresponding to a coupling of the 4d EFT. Upon integrating over
the $S^2$ fibers given by $(u,v)$, (\ref{eq:superp1}) reduces to
the superpotential on the hyperelliptic curve
\begin{equation}\label{eq:superp2}
W_{eff}=\int_{\Sigma'} \,T \wedge R\,.
\end{equation}

The fluxes through all the compact cycles are quantized:
\begin{equation}\label{eq:fluxes1}
\int_{A_i} T = N^R_i -\tau N^{NS}_i \;\;,\;\; \int_{C_i} T = c^R_i
-\tau c^{NS}_i\,,
\end{equation}
$N^R_i\,,N^{NS}_i\,,c^R _i\,, c^{NS}_i\; \in \mathbb Z$. However,
the fluxes through the noncompact cycles can vary continuously
and, in fact, we will argue that they have to diverge. We denote
\begin{equation}\label{eq:fluxes2}
-\int_{B_i} T := \beta_i^R-\tau  \beta_i^{NS}\,.
\end{equation}
These quantities will play the role of running gauge couplings.

Given that the $B$-cycles extend to infinity, and both $R$ and $T$
are differentials of the third kind, we need to regulate their $B$
periods. Following~\cite{csw2} we introduce a cut-off at large
distances $x= \Lambda_0$, replacing $P$ and $\tilde P$ by
$\Lambda_0$ and $\tilde \Lambda_0$. For the noncompact
approximation to be consistent, (\ref{eq:superp2}) has to be
finite in the limit $\Lambda_0 \to \infty$. We write $B_i^r$ for
the regularized version of $B_i$, running from $\tilde \Lambda_0$
to $\Lambda_0$ through the $[a_i^-,a_i^+]$ cut.

The $\Lambda_0$ dependence of $\int_{B_i^r} R$ is most easily
obtained~\cite{intri} by doing a monodromy around infinity
$\Lambda_0 ^{3/2}\to {\rm e}^{2\pi i} \Lambda_0 ^{3/2}$.
\footnote{The exponent is the mass dimension of x: $[x]=3/2$,
which follows from $[S]=3$.} In $\Sigma'$ this corresponds to $
B_i ^r \to B_i ^r+C_p+C_{\tilde P} = B_i^r - 2 \sum_{i=1} ^n
A_i\,, $ giving
\begin{equation}\label{eq:prepcutoff}
\int_{B_i^r}\,R = -\frac{1}{2\pi i} (\sum_{i=1}^n S_i) \,{\rm
log}\, \Lambda_0^3 + \ldots
\end{equation}
where $\ldots$ are single valued contributions. Comparing with
(\ref{eq:Rtoinf}),
\begin{equation}\label{eq:fn}
f_n=-4g_n \sum_{i=1}^n S_i\,.
\end{equation}
From (\ref{eq:prepcutoff}), we see that all the periods have the
same ${\rm log}\, \Lambda_0^3$ dependence.

It was shown in~\cite{intri} that the cutoff dependence of $T$ is
exactly the one needed to cancel the logarithmic divergence from
(\ref{eq:prepcutoff}) and yield a finite cutoff independent
$W_{eff}$:
\begin{equation}\label{eq:renormgeom}
\beta_i^R-\tau \beta_i^{NS}\,=\,\frac{1}{2\pi i}(N_i^R-\tau
N_i^{NS})\,{\rm log}\,(\Lambda_0/\Lambda_i)^3\,.
\end{equation}
The $\beta_i$ where defined in (\ref{eq:fluxes2}) and $\Lambda_i$
are a set of finite energy scales. Therefore (\ref{eq:renormgeom})
may be interpreted as a \emph{geometric renormalization} of
certain bare coupling constants $(\beta_i^R,\,\beta_i^{NS})$. This
is the geometric analog of the RG running of the gauge couplings
(see sections \ref{sec:dualym} and \ref{sec:finite}).

%%%%%%%%%%%%%%%%%%%%%%%%%%%%%%%%%%%%%%%%%%%%%%%%%%%%%%%%%%%%%%%%%%%%%
%%%%%%%%%%%%%%%%%%%%%%%%%%%%%%%%%%%%%%%%%%%%%%%%%%%%%%%%%%%%%%%%%%%%%
%%%%%%%%%%%%%%%%%%%%%%%%%%%%%%%%%%%%%%%%%%%%%%%%%%%%%%%%%%%%%%%%%%%%%
%%%%%%%%%%%%%%%%%%%%%%%%%%%%%%%%%%%%%%%%%%%%%%%%%%%%%%%%%%%%%%%%%%%%%
%%%%%%%%%%%%%%%%%%%%%%%%%%%%%%%%%%%%%%%%%%%%%%%%%%%%%%%%%%%%%%%%%%%%%
\section{Counting vacua on curves with punctures}\label{sec:fluxcurve}

In this section we develop the necessary tools to count
supersymmetric flux vacua on the hyperelliptic curve
(\ref{eq:cy2}). We will show that the index formula $\int {\rm
det}(-R-\omega)$ of ~\cite{ashok} is still valid in our case. This
is a priori not obvious, the main issues being that the curve is
noncompact so many quantities need a regulator and the punctures
contribute extra moduli that have to be included. Furthermore,
having fluxes $\beta_i$ that can vary continuously would
immediately lead to an infinite number of vacua. And finally, we
will have to introduce a tadpole cancellation condition.

Counting supersymmetric flux vacua is equivalent to studying the
geometry of the moduli space $\mathcal M$ of $\Sigma'$. There are
different ways of parametrizing it; while from the EFT it is
natural to work with the $S_i$, in the geometrical side it is more
convenient to use the coefficients $f_k$ of the deformation
$f_{n-1}(x)$. More specifically, we parametrize $\mathcal M$ by
combinations $u_k$ of the $f_k$ ($k=1,\ldots, n$) such that
$$
\frac{\partial R}{ \partial u_k}=\zeta_k\,,
$$
giving directly the basis of holomorphic differentials introduced
in (\ref{eq:zeta}) plus $\zeta_n$. This is an efficient and
symmetric way of taking into account the modulus from the puncture
at P and will simplify our formulas.

$\Sigma$ becomes singular when two branch points coincide; this
leads us to define the discriminant
\begin{equation}\label{eq:discr1}
\Delta(u):= \prod_{a<b} (e_a - e_b)^2
\end{equation}
where $e_a := a_i^{\pm}$. We denote the zero locus by
$\Sigma_\Delta$; the moduli space is therefore
\begin{equation}\label{eq:qmod}
{\mathcal M}=\{(u_k) \in {\mathbb C}^n\} \setminus
\Sigma_\Delta\,.
\end{equation}
$\Sigma_\Delta$ is codimension one in $\mathcal M$ and corresponds
to conifold-like singularities: around two coinciding roots we can
always perform a holomorphic change of variables to rewrite the
curve as
$$
u^2+v^2+y^2-x^2=0\,.
$$
Higher order Argyres-Douglas singularities~\cite{ad} occur when
three or more roots coincide, and will be discussed in sections
\ref{sec:finite} and \ref{sec:examples}.

The moduli space is a special Kahler manifold, with metric
\begin{equation}\label{eq:metric}
G_{i \bar l}= -i \int_{\Sigma'} \zeta_i \wedge \bar \zeta _{\bar
l}
\end{equation}
which can be derived from the Kahler potential
\begin{equation}\label{eq:kahler}
K(u,\bar u)= -i \int R \wedge \bar R\,.
\end{equation}
The covariant derivative is
\begin{equation}\label{eq:covder}
\nabla_i V^j=\partial_i V^j+\Gamma_{\phantom{1}ik}^j
V^k\;\;,\;\;\;\Gamma_{\phantom{1}ik}^j=G^{j \bar l} \partial_i
G_{k \bar l}
\end{equation}
($\partial_i := \partial/\partial u^i$) and the curvature tensor
is
\begin{equation}\label{eq:curv}
R_{i \bar j k \bar l}=G_{i \bar s} \partial_k \Gamma^{\bar
s}_{\phantom{1}\bar j \bar l}\,.
\end{equation}

A displacement in $\mathcal M$ deforms the complex structure of
$\Sigma$, so we expect the holomorphic differentials $\zeta_l$ to
mix with the antiholomorphic ones. It is easy to show that the
covariant derivative of a $(1,0)$ form gives a pure $(0,1)$ form:
\begin{equation}\label{eq:mix}
\nabla_i \zeta_j=c_{ij}^{\phantom{ij}\bar k} \bar \zeta_{\bar
k}\;,\;\,c_{ij}^{\phantom{ij}\bar l} := i G^{k \bar l} \int
\nabla_i \zeta_j \wedge \zeta_k\,,
\end{equation}
and the relation with the curvature is
\begin{equation}\label{eq:cR}
R_{i \bar l j \bar k}=-i\,c_{ijm} c^m_{\phantom{m}\bar k \bar
l}\,.
\end{equation}

%%%%%%%%%%%%%%%%%%%%%%%%%%%%%%%%%%%%%%%%%%%%%%%%%%%%%%%%%%%%%%%%%%%%%
%%%%%%%%%%%%%%%%%%%%%%%%%%%%%%%%%%%%%%%%%%%%%%%%%%%%%%%%%%%%%%%%%%%%%
\subsection{Number of supersymmetric solutions} \label{subsec:susytadpole}

We want to count the vacua that preserve $N=1$ supersymmetry. In
the limit $M_{Pl} \to \infty$, supersymmetric solutions are given
by $\partial_i W_{eff} =0$, where $\partial_i :=
\partial/\partial u_i$. As explained before, this limit
corresponds to taking into account only a neighborhood of the
singularity, so that supergravity effects are negligible.

Solutions to these equations may be viewed in two equivalent ways.
If we want to stabilize at a particular point in the moduli space,
$\partial_i W_{eff}=0$ is an on-shell condition that restricts the
possible values of the fluxes to a subspace. Indeed, since
$\partial_i R$ gives by construction a basis of
$H^{1,\,0}(\Sigma')$,
$$
\partial_i W_{eff}=\int T \wedge \partial_i R = 0 \;\;,\;\;
i=1,\ldots,n
$$
implies that
\begin{equation}\label{eq:Tonshell1}
T\,=\,(N^R-\tau N^{NS})\,\tau_{P,\,\tilde P}+\sum_{i=1}^g
(N^R_i-\tau N^{NS}_i)\,\zeta_i\,=\,\sum_{i=1}^n (N^R_i-\tau
N^{NS}_i)\,\zeta_i\,.
\end{equation}
On the other hand, a holomorphic differential is uniquely
specified by giving its $A$-periods.  Indeed, the $B$-periods are
then functions of the period matrix:
\begin{equation}\label{eq:Tonshell2}
\int_{B_j} T= \sum_{i=1}^n (N^R_i-\tau
N^{NS}_i)\,\int_{B_j}\zeta_i\,.
\end{equation}

The other possible point of view is that we can turn on arbitrary
fluxes through all the cycles; this will lift almost all the
degeneracy of the $N=2$ supersymmetric moduli space, leaving only
some number of $N=1$ supersymmetric vacua. Therefore, if we
specify arbitrarily both the $A$ and $B$ fluxes,
(\ref{eq:Tonshell2}) stabilizes the complex moduli of the curve:
\begin{equation}\label{eq:eom1}
\beta^R_j-\tau \beta^{NS}_j=-\sum_{i=1}^n\,(N^R_i-\tau
N^{NS}_i)\,\int_{B_i} \,\zeta_j\,.
\end{equation}

The ingredient that makes the number of vacua finite in compact
Calabi-Yau manifolds is the tadpole cancellation
condition~\cite{ashok}. There is no such constraint in the
noncompact case, since the flux can go off to infinity. However,
the fluxes cannot be arbitrarily large, because once their
associated energy is of order $M_{Pl}$, the noncompact
approximation breaks down: our local variety will be mixed with
far away cycles in the CY. Therefore, in counting the total number
of vacua, we have to impose by hand a tadpole condition. By
analogy with the compact case~\cite{gkp}, we require that
\begin{equation}\label{eq:tadpole1}
\frac{i}{2 {\rm Im} \tau}\, \int_{\Sigma'}\,T \wedge \bar
T\,=L\,.
\end{equation}
Using the on-shell formula (\ref{eq:Tonshell1}) and recalling
(\ref{eq:metric}), the tadpole condition becomes
\begin{equation}\label{eq:tadpole1a}
0 \le L= \frac{1}{2 {\rm Im} \tau}\,G_{i \bar l} U^i \bar U ^l \le
L_*
\end{equation}
where $U^i := N^R_i -\tau N^{NS}_i$. $L_*$ is the maximum value of
$L$, fixed by data of the compact CY that we choose to embed
(\ref{eq:cy2}).

From (\ref{eq:tadpole1a}), the counting of supersymmetric vacua
may be rephrased in terms of the geometry of $\Sigma$: over each
point $(u^k)$ in moduli space we have a `solid sphere' $U^i(u)$,
with volume $L_*$. Each of these allowed points determines a point
in flux space; the number of such points will give the number of
supersymmetric vacua. Furthermore, (\ref{eq:tadpole1a}) shows why
degeneration limits may produce an infinite number of vacua: if
$G_{i \bar l}$ develops a null direction, the tadpole
condition will not bound the number of flux points. In other
words, from this analysis it is not clear how configurations where
one flux goes to infinity and another goes to minus infinity, in a
correlated way such that $L \ge 0$ stays finite, will be ruled
out. The gauge theory analysis will shed light on this point.

Finally, even with the tadpole condition, the number of solutions
to the equations of motion (\ref{eq:eom1}) with continuous fluxes
$\beta_i$ will be infinite. Fortunately, there is a simple way out
of this problem. Recall that the noncompact hyperelliptic curve
should be considered as part of a compact CY. Instead of
parametrizing the fluxes with arbitrary energy scales $\Lambda_i$,
we take them to be integers. Then (\ref{eq:renormgeom}) will
\emph{fix} the energy scales at particular values, depending on
the fluxes. This approach was also taken in ~\cite{gkp} to study
the consequences of the Klebanov-Strassler solution~\cite{ks} and
leads to the usual exponentially large hierarchies of energy
scales, as we show later.

Now we have all the elements to count vacua on complex curves with
punctures; the derivation of the formula for the density of vacua
continues as in \cite{ashok}: the number of
supersymmetric vacua is given by
$$
N_{vac}(L \le L_*)= \int_0 ^\infty dL \, \theta(L_*-L)\,
\sum_{N_R,N_{NS}}\,\delta(L-\frac{1}{2 {\rm Im} \tau}\,G_{i \bar
l} U^i \bar U ^l) \times
$$
\begin{equation} \label{eq:vacua1}
\times \,\int\big(\prod_{i=1}^n d^2u^i \big)\,\delta(\partial W)
\end{equation}
with
$$
\delta (\partial W) := \prod_l \,\delta(\partial_l
W)\,\delta(\partial_{\bar l} W ^*)\, |\,{\rm det}\, \partial^2
W\,|\,.
$$
Here,
\begin{equation} \label{eq:matrixd2w}
\partial^2 W := \left(%
\begin{array}{cc}
  \partial_l \partial_n W & \partial_l \partial_{\bar n} W^* \\
  \partial_{\bar l} \partial_n W & \partial_{\bar l} \partial_{\bar n} W^* \\
\end{array}%
\right)\,.
\end{equation}
Because of $\delta (\partial W)$, we can replace $\partial_l \to
\nabla_l$ in (\ref{eq:matrixd2w}).

The main simplification in the noncompact case is that, since
$\nabla_l \bar \zeta_{\bar n}=0$, $\nabla_l
\partial_{\bar n} W^*=0$, and then
\begin{equation} \label{eq:det}
|\,{\rm det}\,\partial^2 W\,|={\rm det}\,\partial^2 W=|\,{\rm det}
\nabla_l \partial_n W\,|^2\,.
\end{equation}
Therefore \emph{the number of supersymmetric vacua coincides with
the supersymmetry index}, which is topological and, as we shall
see, much easier to compute. On the contrary, in the compact case,
when gravity is not decoupled, the supersymmetric index gives just
a lower bound to the number of vacua.

The final result is
\begin{equation} \label{eq:Ivacua}
N_{vac}^C(L_*)=\frac{(2\pi L_*)^{2n}}{\pi^n (2n)!}\,\int_{\mathcal
M} \, {\rm det}\,(-R)
\end{equation}
where ${\rm det}\,R := {\rm det}_{\bar s \bar r} \big(R^{\bar
s}_{\phantom{1}\bar r k \bar l} \, du^k \wedge d \bar u ^l
\big)\,$. As expected, this coincides with \cite{ashok} when
$M_{Pl} \to \infty$. The index $C$ is introduced for clarity
reasons, to mean that this is the result from the closed string
side.

%%%%%%%%%%%%%%%%%%%%%%%%%%%%%%%%%%%%%%%%%%%%%%%%%%%%%%%%%%%%%%%%%%%%%
%%%%%%%%%%%%%%%%%%%%%%%%%%%%%%%%%%%%%%%%%%%%%%%%%%%%%%%%%%%%%%%%%%%%%
%%%%%%%%%%%%%%%%%%%%%%%%%%%%%%%%%%%%%%%%%%%%%%%%%%%%%%%%%%%%%%%%%%%%%
%%%%%%%%%%%%%%%%%%%%%%%%%%%%%%%%%%%%%%%%%%%%%%%%%%%%%%%%%%%%%%%%%%%%%
%%%%%%%%%%%%%%%%%%%%%%%%%%%%%%%%%%%%%%%%%%%%%%%%%%%%%%%%%%%%%%%%%%%%%
\section{The dual gauge theory}\label{sec:dualym}

In this section we construct the supersymmetric gauge theory which
is dual to the previous gravity configuration. The analysis will
be done along the lines of the Dijkgraaf-Vafa (DV) correspondence,
based on geometric transitions connecting open and closed
superstrings. However our situation is more general and will
require additional techniques.

Let us first quickly review the DV case, which corresponds to the
flux subspace $N^{NS}_i=0$, $\beta^{R}_n=0$ and
$\beta^{NS}_i=\beta^{NS}_n$ for all $i=1,\ldots, n-1$. The large N
duality between open/closed topological strings was derived
in~\cite{vafa}. The role of the holomorphic matrix model and the
relation to $N=1$ SYM was considered in~\cite{dv, intri, dv2}. On
the other hand, in~\cite{cdsw} the DV relation was derived purely
from the field theory side, using the chiral ring relations and
the Konishi anomaly.

Close to the semiclassical limit $|a_i^+-a_i^-| \ll a_i$, $S_i \to
0$, the geometry (\ref{eq:cy2}) corresponds to a product of $n$
independent deformed conifolds. They are cones over $S^3 \times
S^2$ and, while the $S^2$s are collapsed to zero, the $S^3$s have
finite size as measured by $S_i \ne 0$. In the geometric
transition the $n$ 3-spheres $A_i$ are collapsed and we blow-up
the conifolds at $x=a_i$ by introducing $n$ $\mathbb P^1$'s. Then
the $RR$ fluxes $N_i^R$ will disappear and, instead, we will have
$N_i^R$ D5 branes wrapping the corresponding $\mathbb P^1$s. The
DV correspondence states that the large $N ^R := \sum_{i=1}^n
N_i^R$ limit of the closed string theory on the deformed threefold
is equivalent to the open string theory on the resolved threefold,
with the previous relation between RR fluxes and D5 branes.

$W(x)$ plays the role of a tree-level superpotential for the
chiral superfield $\Phi$ in the $N=2$ vector multiplet of a pure
$U(N^R)$ SYM; this potential breaks $N=2$ to $N=1$. Classically, the
number of vacua is given by the number of ways of choosing $N_i^R$
eigenvalues of $\Phi$ equal to $a_i$, with $\sum_i N_i^R = N^R$.
This breaks $U(N^R) \to \prod_i U(N_i^R)$. $\beta^{NS}_n$ is the
bare gauge coupling of $U(N^R)$, while $c_i^R$ are relative
changes in the $\theta$-angles of the $U(N_i^R)$
factors~\cite{cdsw}. Furthermore, the complex moduli measure
gaugino condensation
\begin{equation} \label{eq:chiralring3}
S_i=-\frac{1}{32 \pi^2} \langle {\rm Tr}\; W_\alpha W^\alpha P_i \rangle
\end{equation}
($P_i$ projects onto $\Phi=a_i$).

%%%%%%%%%%%%%%%%%%%%%%%%%%%%%%%%%%%%%%%%%%%%%%%%%%%%%%%%%%%%%%%%%%%%%
%%%%%%%%%%%%%%%%%%%%%%%%%%%%%%%%%%%%%%%%%%%%%%%%%%%%%%%%%%%%%%%%%%%%%
\subsection{Dualities and geometric transition} \label{subsec:geomtrans}

We return now to the general flux configuration
$(N_i^R,N_i^{NS})$, $(\beta_i^R, \beta_i^{NS})$. Denote $N^R :=
\sum_{i=1}^n \, N_i^R$, $N^{NS} := \sum_{i=1}^n \, N_i^{NS}$ and
$r={\rm gcd}(N^R, N^{NS})$, i.e., $N^R=n_Rr$ and $N^{NS}=n_{NS }r$
with $n_R$ and $n_{NS}$ relatively prime.

Consider first the effect of the geometric transition around the
semiclassical regime. In the open string side we end with $N_i^R$
D5-branes and $N_i^{NS}$ NS5-branes wrapping the i-th $\mathbb
P^1$. The $\beta_i$ do not have a brane analogue since the
$B$-cycles remain 3-cycles; their meaning will become clear later.
Our aim is to find a gauge theory interpretation for these $n$
$(N^R_i,N^{NS}_i)$ 5-brane states. The basic requirement is that
the infrared limit of this configuration shall be given by
composite fields $S_i$ with an effective superpotential
\begin{equation} \label{eq:superp3}
W_{eff}=\sum_{i=1}^n(N_i^R - \tau N^{NS}_i)\,\frac{\partial
\mathcal F}{\partial S_i}-2\pi i \sum_{i=1}^n (\beta_i^R - \tau
\beta^{NS}_i) S_i\,;
\end{equation}
we omitted a $(-1/2\pi i)$ factor as compared to
(\ref{eq:superp2}).

We expect each $(N^R_i,N^{NS}_i)$ 5-brane to
decay to $r_i$ copies of an $(n_i^R,n_i^{NS})$ bound
state~\cite{pol}; here $N^R_i = n_i^R r_i$, $N^{NS}_i = n_i^{NS}
r_i$ with $n_i^R$ and $n_i^{NS}$ coprime. However, the generic
point in flux space will give $n$ different types of bound states
and it is hard to see how this may come from a unique UV gauge
theory. Instead, the straightforward way of getting a gauge theory
is if on each $\mathbb P^1$ we have \emph{the same} type of bound
state. Combining this with the requirement that the sum of fluxes
$(N^R=n_R r, N^{NS}=n_{NS} r)$ remains constant implies that we
will have $r$ copies of the bound state of type $(n_R,n_{NS})$
distributed over all the different $\mathbb P^1$s.

The physical mechanism that may be responsible for this is already
known, namely, eigenvalue tunnelling in matrix models. Consider
what happens when we tune the couplings $a_k$ from (\ref{eq:W1})
so that the $n$ cuts come very close together:
$y^2=x^{2n}+\epsilon$, $\epsilon \to 0$. In this limit, the
process of eigenvalue tunnelling between different cuts becomes
relevant; this will result in RR flux transfer until we end with
the same $(n_R,n_{NS})$ bound states in all the cuts. The
tunnelling is explained by $D5$ branes wrapped around an $S^3$
interpolating between two $S^2$s in the resolved geometry
\cite{dv2}. This object is a domain wall from the EFT point of
view, with tension $\partial \mathcal F /
\partial S_i-\partial \mathcal F / \partial S_j$. After the
tunnelling has taken place, we can tune back the couplings
to their initial values.

We will now start to argue that the previous gauge theory is indeed the dual
to our gravity configuration. The key elements entering into the
argument are $S$-duality (decay to bound states) and moving the
$A_i$ cycles around, which is associated to an $Sp(2n-2,\mathbb
Z)$ symmetry transformation. We work in the deformation side. Denote the deformed threefold defined in (\ref{eq:cy1})
and (\ref{eq:cy2}) by $X_d$; the limit $f_{n-1}(x)=0$ is a singular CY $X_s$ with (generically)
conifold degenerations.

Recall that $S$-duality acts by $SL(2,\mathbb Z)$ transformations
\begin{equation} \label{eq:Sdual}
\left( \begin{array}{c} F_3 \\ H_3\end{array} \right) \to \left(
\begin{array}{cc} a & b \\ c & d \end{array} \right)\,\left(
\begin{array}{c} F_3 \\ H_3\end{array} \right)\;,\;\tau \to
\frac{a\tau+b}{c\tau+d}\;,\;ad-bc=1\,.
\end{equation}
This doesn't change the geometry of the hyperelliptic curve
(off-shell). On the other hand, the curve (\ref{eq:cy2}) has a
symmetry group $Sp(2n-2,\mathbb Z)$ of matrices mixing the
canonical cycles $(A_i\,,\,C_j)$. These transformations are
generated by all the possible interchanges of the roots $a_i^\pm$.
The generators are \cite{theta}
\begin{equation} \label{eq:gens}
J=\left( \begin{array}{cc} 0 & \mathbb I \\ -\mathbb I & 0
\end{array} \right)\;\;,\;\; {\mathcal A}=\left( \begin{array}{cc} (A^t)^{-1} & 0 \\ 0 &
A \end{array} \right)\;\;,\;\; {\mathcal B}=\left(
\begin{array}{cc} \mathbb I & 0 \\ B & \mathbb I
\end{array} \right)\,.
\end{equation}
$A \in GL(n-1,\,\mathbb Z)$ and $B$ is a symmetric matrix with
integer coefficients. Note that $A_1+\ldots+A_n=-C_P$ is invariant
under $Sp(2n-2,\mathbb Z)$ because the loop around infinity
doesn't change under monodromies of the roots.

The first step is to use $S$ duality to set the total NS flux
$N^{NS}=0$ and hence $N^R=r$. The transformation doing this is
\begin{equation} \label{eq:Sdual2}
\left( \begin{array}{c} n_R r \\ n_{NS}r\end{array} \right) \to
\left(
\begin{array}{cc} a & -b \\ -n_{NS} & n_R \end{array} \right)\,\left(
\begin{array}{c} n_R r \\ n_{NS}r\end{array} \right)=\left(\begin{array}{c} r \\ 0\end{array} \right)\,
\end{equation}
for some integers $(a,b)$ solving $an_R-bn_{NS}=1$. We denote with
tildes the transformed quantities after $S$ duality.

Next we set $\tilde N^{NS}_i=0$, $i=1,\ldots, n-1$ with $Sp(2n-2,\mathbb
Z)$ transformations. This is done with the `diagonal'
$SL(2,\mathbb Z)_i \subset Sp(2n-2,\mathbb Z)$ which mix the $A_i$
and $C_i$ cycles only:
\begin{equation} \label{eq:Sp2}
\left( \begin{array}{c} \tilde N_i^{NS} \\ \tilde
c_i^{NS}\end{array} \right) \to \left(
\begin{array}{cc} a_i & b_i \\ c_i & d_i \end{array} \right)\,\left( \begin{array}{c} \tilde N_i^{NS} \\ \tilde
c_i^{NS}\end{array} \right)=\left(\begin{array}{c} 0 \\
\tilde c_i^{\,'NS}\end{array} \right)\,.
\end{equation}
Primes refer to the transformed cycles. Symplectic transformations
act in a complicated way on $A_n$; however, since we already fixed
$N^{NS}=0$ and $A_1+\ldots+A_n$ is a symplectic invariant, we
deduce that the combined application of (\ref{eq:Sdual2}) and
(\ref{eq:Sp2}) fixes \emph{all} $\tilde N_i^{'NS}=0$, $i=1,\ldots,
n$.

\begin{figure}
%\centering
\begin{picture}(0,0)%
\includegraphics[scale=0.8]{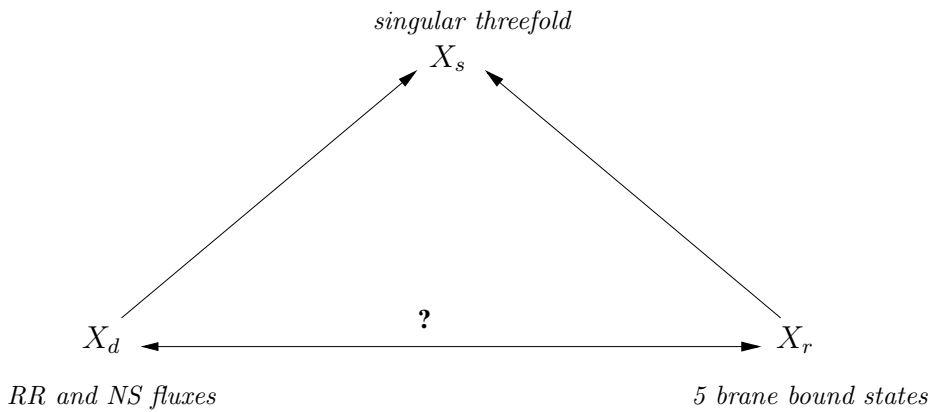}%
\end{picture}%
\setlength{\unitlength}{3100sp}%
\begingroup\makeatletter\ifx\SetFigFont\undefined%
\gdef\SetFigFont#1#2#3#4#5{%
  \reset@font\fontsize{#1}{#2pt}%
  \fontfamily{#3}\fontseries{#4}\fontshape{#5}%
  \selectfont}%
\fi\endgroup%
\begin{picture}(6150,3120)(2176,-5011)
\put(5551,-2311){\makebox(0,0)[lb]{\smash{\SetFigFont{12}{14.4}{\rmdefault}{\mddefault}{\updefault}{\color[rgb]{0,0,0}$X_s$}%
}}}
\put(2776,-4561){\makebox(0,0)[lb]{\smash{\SetFigFont{12}{14.4}{\rmdefault}{\mddefault}{\updefault}{\color[rgb]{0,0,0}$X_d$}%
}}}
\put(8326,-4561){\makebox(0,0)[lb]{\smash{\SetFigFont{12}{14.4}{\rmdefault}{\mddefault}{\updefault}{\color[rgb]{0,0,0}$X_r$}%
}}}
\put(5101,-2011){\makebox(0,0)[lb]{\smash{\SetFigFont{10}{12.0}{\rmdefault}{\mddefault}{\itdefault}{\color[rgb]{0,0,0}singular threefold}%
}}}
\put(7651,-5011){\makebox(0,0)[lb]{\smash{\SetFigFont{10}{12.0}{\rmdefault}{\mddefault}{\itdefault}{\color[rgb]{0,0,0}5 brane bound states}%
}}}
\put(2176,-5011){\makebox(0,0)[lb]{\smash{\SetFigFont{10}{12.0}{\rmdefault}{\mddefault}{\itdefault}{\color[rgb]{0,0,0}RR and NS fluxes}%
}}}
\end{picture}
\caption{\small Geometric transition in the presence RR and NS
fluxes.} \label{fig:trans}
\end{figure}

Summarizing, we have showed how $S \otimes Sp(2n-2,\mathbb Z)$ may
be used to set all the NS fluxes through the $A$ cycles to zero.
The transformed axio-dilaton is $\tilde \tau = (a \tau
-b)/(-n_{NS} \tau+n_R)$; the transformation of $\beta_i$ will be
analyzed shortly. Consider next the effect of the geometric
transition~\cite{dd} $X_d \to X_s \to X_r$ where $X_r$ is the
projective resolution blowing-up each conifold point in $X_s$ to a
$\mathbb P^1$; see Figure \ref{fig:trans}. We end with $r$ copies
of \emph{the same} 5-brane bound state $(n_R,n_{NS})$, wrapping
the $n$ $\mathbb P^1$ s. The gauge theory is then $U(r)\to \prod_i
U(\tilde N_i^{'R})$ where $\tilde N_i^{'R}$ is the number of
$(n_R,n_{NS})$ bound states on the i-th $\mathbb P^1$. This is in
agreement with our previous bound state reasoning in terms of
eigenvalue tunnelling. The 3-cycles $B_i$ don't collapse in the
geometric transition, so in the open string side we still have the
fluxes $(\beta_i^R, \beta_i^{NS})$.

%%%%%%%%%%%%%%%%%%%%%%%%%%%%%%%%%%%%%%%%%%%%%%%%%%%%%%%%%%%%%%%%%%%%%
%%%%%%%%%%%%%%%%%%%%%%%%%%%%%%%%%%%%%%%%%%%%%%%%%%%%%%%%%%%%%%%%%%%%%
\subsection{Properties of the gauge theory} \label{subsec:gaugeproperties}

We don't know how to prove the duality $X_d \longleftrightarrow
X_r$ conjectured in the previous subsection. Although the
introduction of both RR and NS fluxes through the compact cycles
is a natural extension of the Dijkgraaf-Vafa duality, an open
topological string description of $(n_R,n_{NS})$ 5-brane bound
states is not available. Instead, by computing the effective
superpotential for both sides, we shall show that their
predictions agree in the IR limit. As a further check, in section
\ref{sec:examples} we will prove that the gravity and gauge
descriptions have the same number of degrees of freedom even in
strongly coupled regimes, such as Argyres-Douglas singularities.
$(p,q)$ fivebranes wrapping an $S^2$ have also been considered in
the different context of $N=1^*$ SYM~\cite{polstrass}.

Consider how the effective flux superpotential (\ref{eq:superp3})
transforms under the $S \otimes Sp(2n-2,\mathbb Z)$ transformation
given by (\ref{eq:Sdual2}) and (\ref{eq:Sp2}):
$$
\tilde W _{eff}'=\sum_{i=1}^n\,\tilde N_i^{'R}\frac{\partial \mathcal F}{\partial S_i'}-2\pi i \sum_{i=1}^n \left(\frac{\beta_i^{'R}-\tau \beta_i^{'NS}}{n_R-\tau n_{NS}}\right)S_i'\,.
$$
We made explicit the $S$ duality transformation in the second term
to exhibit the fractional dependence on $(n_R-\tau n_{NS})$; apart
from this, $(\tilde N_i^{'R},\,\beta_i^{'R},\,\beta_i^{'NS})$ are
all integers. Rename $\tilde N_i^{'R} \to N_i$ and drop
all the primes:
\begin{equation} \label{eq:superp4}
W_{eff}=\sum_{i=1}^n\, N_i \frac{\partial \mathcal F}{\partial S_i}-2\pi i \sum_{i=1}^n \left(\frac{\beta_i^R-\tau \beta_i^{NS}}{n_R-\tau n_{NS}}\right)S_i\,.
\end{equation}
Here $(N_i,\,\beta_i^R,\,\beta_i^{NS})$ are arbitrary integers
and shouldn't be confused with the original parameters appearing
in (\ref{eq:superp3}).

Let us spell out the holomorphic properties of the gauge theory.
Six dimensional gauge theories based on $(p,q)$ 5-branes were
studied for example in \cite{witten}. The situation here is more
complicated, because the bound states are wrapping $\mathbb P^1$
s, and there is $(\beta_i^R, \beta_i^{NS})$ flux through such
cycles.

Given that we have the same bound states $(n_R,n_{NS})$ in every
$\mathbb P^1$, it is enough to study a single bound state wrapping
a $\mathbb P^1$ and extending in four space-time dimensions. Since
$n_R$ and $n_{NS}$ are relatively prime, the S-duality
transformation (\ref{eq:Sdual2}) maps the bound state to a single
D5 brane. We denote with tildes the variables after the
transformation. The DBI action is~\cite{pol}
$$
S=S_{kin}+S_{CS}
$$
$$
S_{kin}=-\mu_5 \int d^4x\,\int_{S^2} d\Omega_2\,{\rm e}^{-\tilde \Phi} \big[-{\rm det}(\tilde G+\tilde B+F) \big]^{1/2}
$$
\begin{equation} \label{eq:dbi1}
S_{CS}=i\mu_5  \int \big[\tilde C_6+(\tilde B+F)\wedge\tilde C_4+\frac{1}{2}(\tilde B+F)^2\wedge \tilde C_2+\frac{1}{6}(\tilde B+F)^3\,\tilde C_0 \big]\,.
\end{equation}
$F:=2\pi \alpha' F_{ab}$ denotes the $U(1)$ gauge field on the D-brane. Near the geometric transition point, where the $S^2$ shrinks, the holomorphic gauge coupling is given by
\begin{equation} \label{eq:dbi2}
\tilde \tau_{YM}=(2\pi \alpha')^2 \mu_5 \Big(\int_{S^2}\tilde C_2-(\tilde C_0+i{\rm e}^{-\tilde \Phi})\int_{S^2} \tilde B_2 \Big)\,.
\end{equation}

The action for the $(n_R,n_{NS})$ bound state and the properties of its gauge theory follow from (\ref{eq:dbi1}) and S-duality:
$$
\tilde \tau=\tilde C_0+i {\rm e}^{- \tilde \Phi}=\frac{a \tau -b}{-n_{NS} \tau + n_R}\,,
$$
$$
\tilde C_2=a C_2-b B_2\;,\;\,\tilde B_2=-n_{NS}C_2+n_R B_2\,
$$
$$
\tilde G_{ab}=|n_R-n_{NS} \tau|\,G_{ab}\;,\;\,\tilde C_4=C_4
$$
\begin{equation} \label{eq:Stransf}
\tilde B_6-\tilde \tau \tilde C_6=\frac{B_6-\tau C_6}{n_R-\tau n_{NS}}\,.
\end{equation}
Noting that
$$
\int_{S^2} (C_2-\tau B_2)= \beta^R -\tau \beta^{NS}\,,
$$
the gauge coupling becomes
\begin{equation} \label{eq:gaugecoupl}
\tilde \tau_{YM}=\frac{\beta^R -\tau \beta^{NS}}{n_R-\tau n_{NS}}\,,
\end{equation}
where we set $(2\pi \alpha')^2 \mu_5=1$. This coincides exactly
with the fractional holomorphic coupling derived from the flux
side, eq. (\ref{eq:superp4}). Furthermore, once we map the system
of $(p,q)$ 5-branes to D5 branes,  the arguments of \cite{cdsw}
may be applied to this N=1 SYM theory to deduce that the effective
superpotential has precisely the form given in (\ref{eq:superp4}).
Generalizing to the case of $n$ $\mathbb P^1$s, the gauge theory
is $U(r) \to \prod_i U(N_i)$, $\sum_i N_i = r$, and each $U(N_i)$
has a holomorphic coupling
\begin{equation} \label{eq:fractau}
\tau_i := \frac{\beta_i^R-\tau \beta_i^{NS}}{n_R-n_{NS}\tau}\,.
\end{equation}

From our previous construction, it is clear that we didn't fix all the
symplectic symmetries. In particular, we can still perform
monodromies $S_i \to {\rm e}^{2\pi i}\,S_i$ corresponding to $B_i \to
B_i + A_i$. This implies that $\tau_i$ is defined only modulo
$N_i$ or, equivalently,
\begin{equation} \label{eq:valuesbeta}
\beta_i^R=0,\ldots, n_R N_i-1\,;\;\,\beta_i^{NS}=0,\ldots,
n_{NS} N_i-1\,.
\end{equation}
We thus see that the information in the original brane system is
not lost after the $S$-duality $(N^R,N^{NS})\to (r,0)$, but rather
it is encoded in the holomorphic gauge couplings of the new
theory.

It is worth noting that the holomorphic couplings $\tau_i$,
besides being fractional, they are also independent since we can
choose arbitrary integers $\beta_i$. Equivalently from
(\ref{eq:renormgeom}), each $U(N_i)$ factor has an independent
physical scale $\Lambda_i$. This situation is natural from the DBI
action, but it cannot arise as the IR limit of the usual $N=2$
$U(r)$ SYM broken to $N=1$ by the tree level superpotential
$W(\Phi)$. Let us exhibit a simple generalization that may account
for independent $\tau_i$ s. Coming from string theory, we won't
require this UV gauge theory to be renormalizable, so we look for
a modified kinetic term
\begin{equation} \label{eq:modgauge}
\mathcal L_{kin} \sim \int d^2 \theta \, {\rm Tr}\,(W^\alpha
W_\alpha \,f(\Phi))\,.
\end{equation}
If $W(\Phi)=0$, the gauge group is not broken and
$f(\Phi_{class})=\tau_{YM}$ should give a unique gauge coupling.
On the other hand, when we turn on the superpotential, the basic
property of $f(\Phi)$ is that it should be equal to $\tau_i$ on
the subspace $\Phi=a_i$. The matrix function that does this is
simply constructed from the idempotents of the classical chiral
ring:
\begin{equation} \label{eq:idemp1}
E_i(\Phi)=\frac{\prod_{j \ne i}(\Phi-a_j \mathbb I)}{\prod_{j \ne
i}(a_i-a_j)}\,,
\end{equation}
which satisfy $E_i(a_j)=\delta_{ij}$. Then we may define
\begin{equation} \label{eq:modgauge2}
f(\Phi) := \sum_{i=1}^n \tau_i E_i(\Phi)\,.
\end{equation}
The nonrenormalizable gauge theory (\ref{eq:modgauge}) with this
choice of $f(\Phi)$ gives independent gauge couplings in the
infrared.

Another striking property of this brane system is the appearance
of noncommutative dipoles in the UV. This is due to the NS fluxes
through the $\mathbb P^1$ s. Such dipole deformations of the gauge
theory have been recently considered in~\cite{dipole} for
geometric transitions based on D5 branes. It would be interesting
to try to extend this analysis to the case of $(n_R, n_{NS})$
5-branes, although the supergravity description might be much more
involved.

To summarize, using $S \otimes Sp(2n-2,\mathbb Z)$ in this section
we mapped a general flux configuration to a gauge theory, after
the geometric transition. All the flux parameters have a natural
gauge interpretation; in particular the fluxes $(\beta_i^R,
\beta_i^{NS})$ through the 3-cycles, which don't collapse after
the transition, don't contribute brane degrees of freedom. They
combine in a nontrivial way to determine the holomorphic gauge
couplings of the different gauge factors.

%%%%%%%%%%%%%%%%%%%%%%%%%%%%%%%%%%%%%%%%%%%%%%%%%%%%%%%%%%%%%%%%%%%%%
%%%%%%%%%%%%%%%%%%%%%%%%%%%%%%%%%%%%%%%%%%%%%%%%%%%%%%%%%%%%%%%%%%%%%
%%%%%%%%%%%%%%%%%%%%%%%%%%%%%%%%%%%%%%%%%%%%%%%%%%%%%%%%%%%%%%%%%%%%%
%%%%%%%%%%%%%%%%%%%%%%%%%%%%%%%%%%%%%%%%%%%%%%%%%%%%%%%%%%%%%%%%%%%%%
%%%%%%%%%%%%%%%%%%%%%%%%%%%%%%%%%%%%%%%%%%%%%%%%%%%%%%%%%%%%%%%%%%%%%
\section{Finiteness of vacua in the dual gauge side}\label{sec:finite}

The purpose of constructing a dual gauge theory to count flux
vacua is that in such field theories the number of vacua is always
finite. The geometric transition preserves this number. In the
present section we show from the gauge theory side that $N_{vac}$
is indeed finite.

%%%%%%%%%%%%%%%%%%%%%%%%%%%%%%%%%%%%%%%%%%%%%%%%%%%%%%%%%%%%%%%%%%%%%
%%%%%%%%%%%%%%%%%%%%%%%%%%%%%%%%%%%%%%%%%%%%%%%%%%%%%%%%%%%%%%%%%%%%%
\subsection{Proof of the finiteness of $N_{vac}$} \label{subsec:countdv}

We begin by showing that the number of supersymmetric gauge vacua,
i.e., solutions to $\partial W_{eff}/\partial S_i$ from equation
(\ref{eq:superp4}), is finite. As discussed before, this is based
on the tadpole constraint
\begin{equation} \label{eq:gaugetad}
L=\sum_{i=1}^n N_i \tilde \beta_i^{NS}\,.
\end{equation}
Here $\tilde \beta_i^{NS}=(n_R \beta_i^{NS}-n_{NS} \beta_i^R)$;
also recall that $N_i:=\tilde N_i^{'R}$,
$\beta_i^R:=\beta_i^{'R}$, $\beta_i^{NS}:=\beta_i^{'NS}$.

We have to sum over all choices of fluxes satisfying
(\ref{eq:gaugetad}). Here we run into the main obstacle. The
reason why this could in principle diverge is that there may be
flux configurations such that two terms in $L$ grow in a
correlated way to plus and minus infinity respectively, but
keeping $L$ finite and positive. This would give an infinite
number of allowed flux points (and hence supersymmetric vacua).

This is the point where having a gauge theory based on the
geometry (\ref{eq:cy2}) proves useful. In the gauge theory,
$W_{eff}$ is holomorphic in the couplings $a_k$, so the number of
solutions to the equations $\partial W_{eff}/\partial f_i = 0$ is
invariant under smooth changes of the parameters, being protected
by holomorphy.\footnote{Since off-shell the $f_i$ don't depend on
$a_k$, it is more convenient to take derivatives w.r.t. $f_i$ and
not $S_i$.}  An equivalent statement is that the number of vacua
coincides with the dimension of the chiral ring of the theory, and
such a quantity is independent of the gauge couplings. This
topological behavior was already encountered in the gravity side,
when we showed (section \ref{subsec:susytadpole}) that the number
of supersymmetric vacua coincides with the supersymmetric index.

We now argue, from a variation of the $a_k$, that each term in
$L$ is in fact positive even around singularities. The discriminant locus consists
of generic conifold points and higher codimension AD singularities. The later cannot
be neglected because they have a higher `weight' in the counting of degrees of
freedom, as measured by ${\rm det} (R)$. Both situations will be exemplified in section \ref{sec:examples}.

Consider a point in moduli space
$\mathcal M$ corresponding to the semiclassical limit. This is
just the origin $S_i \to 0$ of $\mathcal M$. In this case the
geometry is a product of independent conifold-like configurations.
The effective superpotential follows from (\ref{eq:superp4}) using
monodromy arguments~\cite{intri}:
\begin{equation} \label{eq:superpconif}
W_{eff}=\sum_{i=1}^n\, N_i S_i \big({\rm
log}(\frac{\Lambda_0^3}{S_i})+1 \big)-2\pi i \sum_{i=1}^n
\left(\frac{\beta_i^R-\tau \beta_i^{NS}}{n_R-\tau
n_{NS}}\right)S_i\,.
\end{equation}
Denoting $\theta_i/2\pi:= {\rm Re}(\tau_i)$ and $1/g_i^2:={\rm Im}(\tau_i)$, the supersymmetric vacua may be written as
\begin{equation}
\label{eq:conif1} S_i={\rm exp}(-i \theta_i/N_i)\,{\rm
exp}(-2\pi /g_i ^2 N_i)\,\Lambda_0^3= {\rm exp}(-i
\theta_i/N_i)\,\Lambda_i^3\,.
\end{equation}

Then counting vacua in the neighborhood of the conifold limit
implies summing over fluxes giving $0 \le |S_i| \le
(\Lambda_i^{\phantom{1}f})^3 $.
\footnote{$ (\Lambda_i^{\phantom{1}f})^3$ is some final energy
scale associated to $U(N_i)$.} Clearly this requires ${\rm{sign}}(n_R \beta_i^{NS}-n_{NS}
\beta_i^R)={\rm{sign}}(N_i^{R})$, to avoid vacua exponentially far
away from the origin. We therefore see that the number of vacua
around the semiclassical point is finite because each term in $L$
is \emph{separately} positive. Without loss of generality, we can
just take all the fluxes to be positive.

The holomorphic dependence of $W_{eff}$ on $a_k$ implies that this
is true for the whole moduli space. Indeed, every point in moduli
space can be connected to the semiclassical limit by such a
variation of couplings. Of course, strongly coupled limits may
have quite complicated superpotentials, but we are interested in
the number of vacua, which is a topological invariant.

For concreteness, we show this for $n=2$. The hyperelliptic curve
is
\begin{equation}
\label{eq:curve1} y^2=(x^2+g_1x+g_0)^2+f_2x+f_1\,.
\end{equation}
We only need to worry about singularities in $\mathcal M$ since it
is known that $N_{vac}$ is finite around smooth points. There are
two types; the codimension one singularities are conifolds, and
correspond to the semiclassical regime where we showed the
finiteness of $N_{vac}$. There is also a codimension two $A_2$
singularity. It corresponds to the singular limit of $y$:
\begin{equation}
\label{eq:curvsing} y^2=(x^3-\delta u x - \delta
v)(x-1)\;\;;\;\;\delta u\,,\,\;\delta v \to 0\,.
\end{equation}
Three roots coincide at $x=0$ giving two vanishing intersecting
cycles, while the last one is fixed at $x=1$. Comparing to
(\ref{eq:curve1}), we find the `double scaling' limit
\begin{equation}
\label{eq:doublelimit1} f_1=\delta v - (\frac{1}{8}+\frac{\delta
u}{2})^2\;,\;\;f_2=-\frac{1}{8}+\frac{\delta u}{2}-\delta v\,,
\end{equation}
and, for the couplings,
\begin{equation}
\label{eq:doublelimit2}
g_1=-\frac{1}{2}\;\;,\;\;g_0=-(\frac{1}{8}+\frac{\delta u}{2})\,.
\end{equation}

\begin{figure}[tbf]
\centering
\begin{picture}(0,0)%
\includegraphics[scale=0.5]{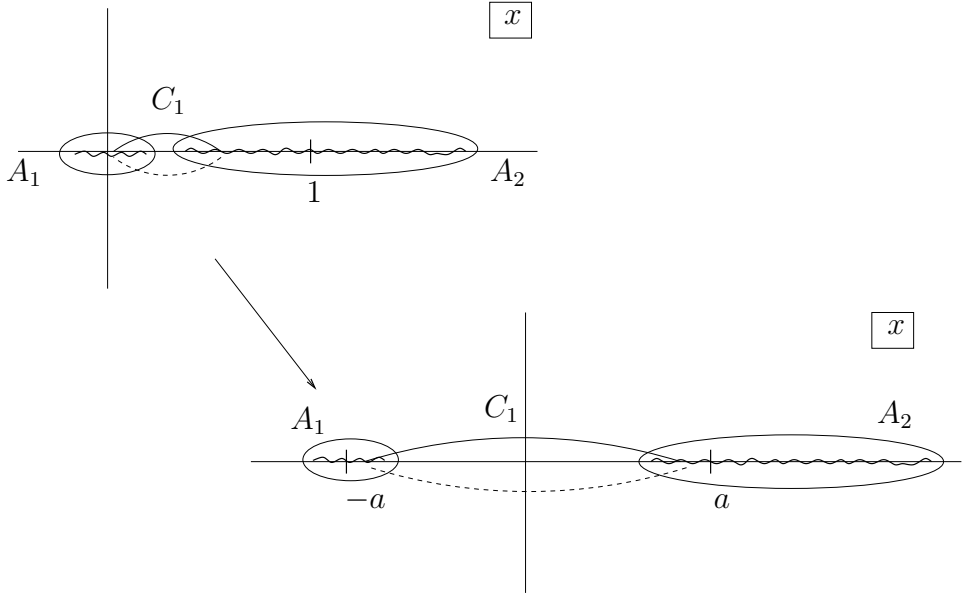}%
\end{picture}%
\setlength{\unitlength}{2000sp}%
\begingroup\makeatletter\ifx\SetFigFont\undefined%
\gdef\SetFigFont#1#2#3#4#5{%
  \reset@font\fontsize{#1}{#2pt}%
  \fontfamily{#3}\fontseries{#4}\fontshape{#5}%
  \selectfont}%
\fi\endgroup%
\begin{picture}(12012,7449)(676,-7423)
\put(4400,-2561){\makebox(0,0)[lb]{\smash{\SetFigFont{12}{14.4}{\rmdefault}{\mddefault}{\updefault}{\color[rgb]{0,0,0}$1$}%
}}}
\put(4201,-5386){\makebox(0,0)[lb]{\smash{\SetFigFont{12}{14.4}{\rmdefault}{\mddefault}{\updefault}{\color[rgb]{0,0,0}$A_1$}%
}}}
\put(11476,-5311){\makebox(0,0)[lb]{\smash{\SetFigFont{12}{14.4}{\rmdefault}{\mddefault}{\updefault}{\color[rgb]{0,0,0}$A_2$}%
}}}
\put(2476,-1411){\makebox(0,0)[lb]{\smash{\SetFigFont{12}{14.4}{\rmdefault}{\mddefault}{\updefault}{\color[rgb]{0,0,0}$C_1$}%
}}}
\put(6601,-5236){\makebox(0,0)[lb]{\smash{\SetFigFont{12}{14.4}{\rmdefault}{\mddefault}{\updefault}{\color[rgb]{0,0,0}$C_1$}%
}}}
\put(6850,-360){\makebox(0,0)[lb]{\smash{\SetFigFont{12}{14.4}{\rmdefault}{\mddefault}{\updefault}{\color[rgb]{0,0,0}$x$}%
}}}
\put(11601,-4200){\makebox(0,0)[lb]{\smash{\SetFigFont{12}{14.4}{\rmdefault}{\mddefault}{\updefault}{\color[rgb]{0,0,0}$x$}%
}}}
\put(676,-2311){\makebox(0,0)[lb]{\smash{\SetFigFont{12}{14.4}{\rmdefault}{\mddefault}{\updefault}{\color[rgb]{0,0,0}$A_1$}%
}}}
\put(6676,-2311){\makebox(0,0)[lb]{\smash{\SetFigFont{12}{14.4}{\rmdefault}{\mddefault}{\updefault}{\color[rgb]{0,0,0}$A_2$}%
}}}
\put(4876,-6361){\makebox(0,0)[lb]{\smash{\SetFigFont{12}{14.4}{\rmdefault}{\mddefault}{\updefault}{\color[rgb]{0,0,0}$-a$}%
}}}
\put(9451,-6361){\makebox(0,0)[lb]{\smash{\SetFigFont{12}{14.4}{\rmdefault}{\mddefault}{\updefault}{\color[rgb]{0,0,0}$a$}%
}}}
\end{picture}
\caption{\small Holomorphic change of couplings that connects the
AD point and the semiclassical limit.} \label{fig:changeg}
\end{figure}

To connect this to the semiclassical point, vary the couplings
$g_i$ from their previous double-scaled values to $g_i \gg f_i$,
while keeping the $f_i$ fixed at (\ref{eq:doublelimit1}). Clearly,
at the new point in $\mathcal M$ the semiclassical approximation
is valid. This process is depicted in Figure \ref{fig:changeg}.

Therefore we have shown that any point in $\mathcal M$ can be
connected to the conifold limit by a smooth variation of the
$a_k$. In other words, the gauge theory tells us how to do, on
every point in moduli space, a change of variables $S_i(a_k) \to
S_i(\tilde a _k)$ such that: (i) each term in $L$ is explicitly
positive and (ii) the number of supersymmetric vacua doesn't
change. Furthermore, since we can work in a regime $f_i \to 0$ by
tuning $a_i \gg f_i$, we can always do power-series expansions and
hence the change of variables is continuous. This maps compact
regions to compact regions, assuring that the number of vacua
doesn't diverge.

The meaning of this transformation becomes transparent if we consider the chiral ring.
It is generated by idempotents and nilpotents~\cite{ring}. If we move around the moduli
space $S_i$ by changing the couplings until we encounter a singularity, the result on the
chiral ring is that some idempotents become nilpotents. The total number of generators is
conserved in the process.

%%%%%%%%%%%%%%%%%%%%%%%%%%%%%%%%%%%%%%%%%%%%%%%%%%%%%%%%%%%%%%%%%%%%%
%%%%%%%%%%%%%%%%%%%%%%%%%%%%%%%%%%%%%%%%%%%%%%%%%%%%%%%%%%%%%%%%%%%%%
\subsection{Formula for $N_{vac}(L_*)$}\label{subsec:formulaNvac}

In order to compare with the gravity side result
(\ref{eq:Ivacua}), we next compute the number of supersymmetric
gauge vacua around an arbitrary point in $\mathcal M$. As argued
before, holomorphy implies that we can as well compute it around
the semiclassical limit.

Because of the monodromies leading to (\ref{eq:valuesbeta}), at
fixed $N_i$, the number of vacua is
\begin{equation} \label{eq:nvac1}
N_{vac}(\{N_i\})=(n_R n_{NS})^n \prod_{i=1}^n N_i^2\,;
\end{equation}
the $N_i$ satisfy $\sum_i N_i=r$. This is quite different to the result
from a standard $N=1$ SYM, $\prod_i N_i$. Eq. (\ref{eq:Ivacua})
includes an integration over a region in moduli space. We need to
specify the analogous condition in the gauge side. It is
associated to the RG flow of the gauge theory from the cutoff
$\Lambda_0$ up to some IR energy scale $\Lambda^f$. For
concreteness, we compute $N_{vac}$ for the simplest case, namely
when each $U(N_i)$ flows up to a scale
$\Lambda_i^{\phantom{1}f}$. In other words, we assume that we are
integrating on disks $0 \le |S_i| \le (\Lambda_i^{\phantom{1}f})^3
$.

The renormalization of gauge couplings
(\ref{eq:renormgeom}) applied to the case (\ref{eq:superp4}) gives
\begin{equation} \label{eq:renormgeom2}
\frac{\tilde
\beta_i^{NS}}{n_R^2+n_{NS}^2}=\frac{1}{2\pi}
N_i\,{\rm log}\big(\frac{\Lambda_0}{\Lambda_i} \big)^3\,.
\end{equation}
Here we set, for simplicity, $C_0=0$, $g_s=1$. This is possible because
in the noncompact model the axio-dilaton is fixed and behaves as a coupling;
therefore $N_{vac}$ cannot depend on it. Since we are summing the degrees of freedom with $0 \le \Lambda_i
\le \Lambda_i^{\phantom{1}f}$, (\ref{eq:renormgeom2}) implies
\begin{equation} \label{eq:renormgeom3}
\tilde \beta_i^{NS} \ge \frac{1}{2\pi}(n_R^2+n_{NS}^2)
N_i\,{\rm log}\big(\frac{\Lambda_0}{\Lambda_i^{\phantom{1}f}}
\big)^3\,.
\end{equation}
Replacing in the gauge tadpole condition (\ref{eq:gaugetad}),
\begin{equation} \label{eq:gaugetad2}
(n_R^2+ n_{NS}^2)\,\sum_{i=1}^n\,N_i^2 \,{\rm log}
\big( \frac{\Lambda_0}{\Lambda_i^{\phantom{1}f}}\big)^3\,\le 2\pi\,L\,.
\end{equation}

Once we fix arbitrary $(N_i)$, the dual fluxes $(\tilde
\beta_i^{NS})$ are integers satisfying the diophantine equation
(\ref{eq:gaugetad}). This has solutions iff ${\rm gcd}(N_i)|L$;
the number of integer solutions is of course infinite, but we
argued that ${\rm sign} (N_i)={\rm sign} (\tilde \beta_i^{NS})$.
So we take the fluxes to be positive, and multiply the number of
vacua by $2^n$. The number of positive solutions to the tadpole
constraint will be denoted by $b_+(\{N_i\})$. For large $L$, this
number is typically of order 1.

Combining all the previous elements, the total number of
supersymmetric vacua is
$$
N_{vac}(L_*;\Lambda^f)=2^n\,{\sum_{L=0}^{L_*}}\;\sum_{n_R,\,n_{NS}\,coprime}\;\,(n_R
n_{NS})^n\,\sum_{\{N_i\}:\,gcd(N_i)\,|L}\,
[\prod_{i=1}^n\,N_i^2]\; \times
$$
\begin{equation} \label{eq:gaugeNfinal}
\times \;b_+(\{N_i\})\,\cdot\,T(N_i; n_R,\,n_{NS})\,.
\end{equation}
The notation here is the following. The sum on $(n_R,n_{NS})$ is
over coprime integers. The sum on $(N_i)$ should be done over
inequivalent fluxes with respect to the residual symplectic
transformations; indeed, some generators in (\ref{eq:gens}) were
not fixed by the mapping to the region
$(N_i^R,N_i^{NS})\,\to\,(N_i^R,0)$. Also, recall that
$b_+(\{N_i\})$ is the number of positive solutions to the
diophantine equation (\ref{eq:gaugetad}); for large $L_*$, it will
give subleading contributions so, to a good approximation, we may
set $b_+ \sim 1$. Lastly, $T(N_i; n_R,\,n_{NS})$ specifies the
region in flux space over which we are summing vacua. For
instance, if we integrate on disks of radius
$(\Lambda_i^{\phantom{1}f})^3$, (\ref{eq:gaugetad2}) gives the
Heaviside function
\begin{equation} \label{eq:Tdisk}
T(N_i; n_R,\,n_{NS})=\Theta \Big(2\pi\,L- (n_R^2+
n_{NS}^2) \sum_{i=1}^n\,{\rm log} \big(
\frac{\Lambda_0}{\Lambda_i^{\phantom{1}f}}\big)^3\,N_i^2\Big)\,.
\end{equation}

%%%%%%%%%%%%%%%%%%%%%%%%%%%%%%%%%%%%%%%%%%%%%%%%%%%%%%%%%%%%%%%%%%%%%
%%%%%%%%%%%%%%%%%%%%%%%%%%%%%%%%%%%%%%%%%%%%%%%%%%%%%%%%%%%%%%%%%%%%%
%%%%%%%%%%%%%%%%%%%%%%%%%%%%%%%%%%%%%%%%%%%%%%%%%%%%%%%%%%%%%%%%%%%%%
%%%%%%%%%%%%%%%%%%%%%%%%%%%%%%%%%%%%%%%%%%%%%%%%%%%%%%%%%%%%%%%%%%%%%
\section{Examples} \label{sec:examples}

In this section we compare the formulas (\ref{eq:Ivacua}) and
(\ref{eq:gaugeNfinal}) for $N_{vac}$ in the gravity and gauge
side, respectively. This is done for the conifold and
Argyres-Douglas degenerations.

%%%%%%%%%%%%%%%%%%%%%%%%%%%%%%%%%%%%%%%%%%%%%%%%%%%%%%%%%%%%%%%%%%%%%
%%%%%%%%%%%%%%%%%%%%%%%%%%%%%%%%%%%%%%%%%%%%%%%%%%%%%%%%%%%%%%%%%%%%%
\subsection{Example 1: the conifold}\label{subsec:conifold}

\vskip 5mm
\emph{Gravity side}
\vskip 5mm

We start by considering the case of a single deformed conifold in
the closed string side. The total number of vacua for the conifold
has been computed in \cite{denef} in the context of F-theory
compactifications. Here we quickly summarize the result for fixed
axio-dilaton.

There is only one compact cycle (A), and a dual noncompact one (B). From monodromy
arguments,
$$
\int_A \Omega=z\;,\;\,\int_B \Omega=\frac{1}{2 \pi i}\,z \,{\rm log} \left(\frac{\mu}{z} \right)+\ldots\,.
$$
$z$ is the complex modulus (here we don't use $S$ to make clear the distinction between
the gravity and gauge side) and $\mu$ is a constant added for dimensional reasons. It
depends on the cutoff necessary to regulate the B-integral. Further, the dots refer to
analytic terms in $z$.

Replacing in (\ref{eq:kahler}) and then in (\ref{eq:curv}),
$$
G_{z \bar z} \approx c\, {\rm log} \left(\mu^2/|z|^2 \right)\;,\;\,R^{z}_{\phantom{1}zzz}=-\frac{1}{|z|^2\, \big({\rm log}\,\mu^2/|z|^2\big)^2}\,.
$$
For $z \to 0$, $G \ll R$ and hence ${\rm det}(-R-\omega) \approx {\rm det} (-R)$, in agreement
with the deduced result (\ref{eq:Ivacua}). Integrating on $0 \le |z| \le R$, the number
of supersymmetric vacua for fixed axio-dilation is
\begin{equation} \label{eq:Nconif1}
N^C_{vac}(L_*)=\frac{2\pi^2\,L_*^2}{{\rm log}\,\frac{\mu}{R}}\,.
\end{equation}
The superindex $C$ reminds us that this is the result from the
closed string side.

\vskip 5mm
\noindent \emph{Gauge theory side}
\vskip 5mm

Next we calculate in detail the result from (\ref{eq:gaugeNfinal}). From the gauge theory side, the conifold corresponds to the
semiclassical limit of the superpotential with $n=1$: $W'(x)=x$
and from (\ref{eq:fn}), $f_{n-1}(x)=f_1=-4 S$, where we are
setting $g_n=1$. There are $N$ vacua satisfying
$$
|S|={\rm e}^{-2\pi /g^2 N}\,\Lambda_0^3 := \Lambda^3
$$
and we have to compute the number of vacua with $|S| \le
\Lambda_f^3$ for some final energy scale $\Lambda_f$. From the
running of the gauge coupling,
$$
\tilde \beta^{NS} \ge \frac{1}{2\pi}\,(n_R^2+
n_{NS}^2)\,N\,{\rm
log}\,(\frac{\Lambda_0}{\Lambda_f})^3\,.
$$
The meaning of this formula is that the gauge theory analogue of
integrating a given modulus on a disk is the RG flow of the gauge
coupling from the UV cutoff up to a final energy scale given by
the radius of the disk.

The number of vacua for given L is then given by
\begin{equation} \label{eq:Nconif2}
N^O_{vac}(L)=2\sum_{N|L}\,N^2\,\sum_{n_R,\,n_{NS}\,coprime}\;n_R
n_{NS}\;\Theta\big(\frac{2\pi L}{{\rm
log}\,(\Lambda_0/\Lambda_f)^3 N^2}\,-\,(n_R^2+ n_{NS}^2)\big)\,;
\end{equation}
we multiply by 2 since we are considering only $N \ge 0$. The superindex
$O$ refers to the open string side.

The gravity result ${\rm det}(-R)$ arises in the continuum limit
$L_* \gg 1$. Therefore we need to estimate the asymptotic behavior of
$\sum_{L=0}^{L_*} N_{vac}^O (L)$. We did this with a C++ program \footnote{We thank S.
Lukic for help with this.} that adds coprime numbers (modulo permutations)
inside a disk of radius
$$
\frac{2\pi L}{{\rm
log}\,(\Lambda_0/\Lambda_f)^3 N^2}
$$
and then sums over all the divisors of $L$, according to (\ref{eq:Nconif2}).

\begin{figure}[tf]
%\centering
\begin{picture}(0,200)%
\includegraphics{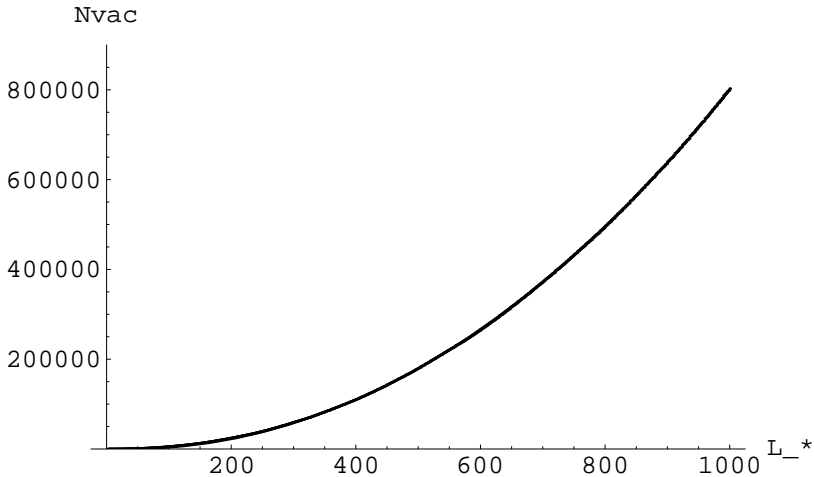}%
\end{picture}%
\caption{\small Plot of $N_{vac}(L_*)$ for the conifold, showing
both the gravity and gauge side predictions, which agree almost
exactly. We chose a scale ${\rm
log}\,(\Lambda_0/\Lambda_f)^3=8\pi$ to simplify the results.}
\label{fig:conif}
\end{figure}

Fitting the numerical predictions of ${\rm log} \, N_{vac}(L_*)$ for $L_*=1000$, we deduce
an asymptotic dependence ${\rm log} N_{vac}(L_*) \approx 2.017 \,{\rm log}\,(L_*)$. To leading
order we find
\begin{equation} \label{eq:Nconif3}
N_{vac}(L_*)= \frac{8 \pi}{{\rm log}\,(\Lambda_0/\Lambda_f)^3}
\big(0.7852\, L_*^{2.017}- 12.370\, L_* \,{\rm log} \,L_* \big)\,.
\end{equation}
The numerical results and the fit are shown in Figure
\ref{fig:conif}. We don't completely understand the subleading
corrections to the gravity result. Even though we fit the
numerical formula with $L_*\,{\rm log}\,L_*$, the power of $L_*$
could be smaller.

Let us compare (\ref{eq:Nconif1}) and (\ref{eq:Nconif3}); we
naturally identify $R := \Lambda_f ^3$ and $\mu := \Lambda_0^3$
and both results match very well. The power $2.017$ is a good
approximation to the gravity result $L_*^2$. It turns out to be
related to properties of the divisor functions $\sigma_k(n)$. The
agreement is nontrivial, involving very different concepts in the
gauge and gravity side. The crucial ingredients from the gauge
side are the running of the gauge coupling and the correct tadpole
condition. In other words, the gravity side with general fluxes
has the same number of degrees of freedom as the SYM theory
described in section \ref{sec:dualym}.

%%%%%%%%%%%%%%%%%%%%%%%%%%%%%%%%%%%%%%%%%%%%%%%%%%%%%%%%%%%%%%%%%%%%%
%%%%%%%%%%%%%%%%%%%%%%%%%%%%%%%%%%%%%%%%%%%%%%%%%%%%%%%%%%%%%%%%%%%%%
\subsection{Example 2: Argyres-Douglas singularities}\label{subsec:adgravity}

Next we analyze some aspects of two-parameter models which arise
from $n=2$ superpotentials:
\begin{equation} \label{eq:n2}
y^2=(x^2+g_1 x+g_0)^2+f_2x+f_1\,.
\end{equation}
The novel phenomenon for $n \ge 2$ is the appearance of
Argyres-Douglas points, when three or more roots coincide; see
(\ref{eq:curvsing}). When intersecting cycles vanish
simultaneously nonlocal dyons become massless. The physics is
radically different to that of the conifold, giving rise to an
interacting SCFT \cite{ad}.

Unfortunately, the complications of the model forbid a
straightforward analysis similar to the one done in the previous
subsection. From the gravity side, the discriminant locus is a
knot-like complex curve \cite{ad} with self-intersections;
integrating over all the moduli space to get the total number of
vacua is hence quite involved. On the other hand, in the gauge
theory, the combinatorics present in formula
(\ref{eq:gaugeNfinal}) are equally complicated. Therefore we will
only study the vicinity of the AD point and we will show that the
number of vacua obtained from ${\rm det} \,R$ has the expected
gauge theory scaling behavior.

\vskip 5mm \noindent \emph{Gravity side}\footnote{Done in
collaboration with F. Denef and B. Florea.} \vskip 5mm

The dynamics around the AD point is controlled by the small
complex curve
\begin{equation} \label{eq:adcurve2}
w^2=x^3-\delta u x -\delta v\,.
\end{equation}
The discriminant locus is
\begin{equation}\label{eq:discr3}
\Delta\,=\,4(\delta u)^3-27(\delta v)^2=0\,,
\end{equation}
which is not smooth; indeed
$$
\Delta=0\;\;,\;\;\partial_{\delta u} \Delta=\partial_{\delta v}
\Delta=0
$$
has solution $(\delta u=0\;,\;\delta v=0)$. This is the
Argyres-Douglas point~\cite{ad}. Usual monodromy arguments used to
construct the periods cannot be applied now, since the
self-intersection is not normal. Therefore we need to blow-up
(\ref{eq:discr3}). The general procedure is described
in~\cite{arnold} and has been recently applied to our present
situation in~\cite{eguchi}.

The normal-crossing variables close to the AD point turn out to be
\begin{equation}\label{eq:localcoord}
\Delta := \frac{(\delta u)^3}{(\delta
v)^2}-\frac{27}{4}\;\;,\;\;\eta := \frac{\delta v}{\delta u}\,.
\end{equation}
The original discriminant locus corresponds to $\Delta=0$; $\eta$
is the scaling variable in the SCFT. By rescaling $x=\eta \tilde
x$, $w=\eta^{3/2} \tilde w$ the dependence on $\eta$ disappears;
the dependence on $\Delta$ follows from the usual monodromy
$\Delta \to {\rm e}^{2\pi i}\,\Delta$.

We do a symplectic transformation so that the small periods are
$(S_1,\,\frac{\partial {\mathcal F}}{\partial S_1})$ and the large
ones are $(S_2,\,\frac{\partial {\mathcal F}}{\partial S_2})$. The
dependence on $\Delta$ and $\eta$ is
\begin{equation}\label{eq:AD periods}
S_1 \propto \eta^{5/2}\,\Delta\;,\,\,\frac{\partial \mathcal
F}{\partial S_1} \propto \eta^{5/2}\,\Delta\,{\rm log}\,\Delta\,.
\end{equation}
The large periods are analytic in $\Delta$ and $\eta$.

Replacing these expressions in (\ref{eq:metric}) and
(\ref{eq:curv}), the density of vacua (\ref{eq:Ivacua}) around the
AD point is
\begin{equation} \label{eq:IAD}
dN_{vac}\,\propto\,\frac{L_*^4\; d^2\,\Delta\,d^2\eta}{|\eta|\,|\Delta|\,^2\,
({\rm log}\,|\Delta|)^3}\,.
\end{equation}
We see that the density of vacua is integrable on a disk around
$(\Delta,\eta)=(0,0)$. In particular, integrating on $0 \le |\Delta| \le \Lambda_f^3$ gives
a total number of vacua
\begin{equation} \label{eq:IAD2}
N_{vac}^C(L_*, \Lambda_f)=\frac{2 \pi^2 k\,L_*^4}{({\rm
log}\,(\Lambda_0/\Lambda_f)^3)^2}\,.
\end{equation}
This proves that the number of vacua around the AD singularity is
finite. The constant $k$ depends on analytic data from the long
cycles; in general it cannot be computed using monodromy
arguments.

The result (\ref{eq:IAD}) is of the general form encountered in
the analysis of different singularities in \cite{eguchi}
\begin{equation} \label{eq:eguchi}
dN_{vac} \sim \, \frac{dz\,d \bar z}{|z|^2({\rm log}|z|)^p}
\end{equation}
where $z=0$ denotes de discriminant locus (in normal crossing
variables). We will now justify this behavior from the field
theory point of view.

\vskip 5mm
\noindent  \emph{Gauge side}
\vskip 5mm

This example is quite interesting, since we have to use the map
connecting the strongly coupled AD point to the semiclassical
regime.

The procedure was described in section \ref{sec:finite}. We vary
$g_k$ from (\ref{eq:doublelimit2}) to $W'(x)=x^2-a^2$, while
keeping $f_i$ fixed at (\ref{eq:doublelimit1}). The condition that
we end in the semiclassical regime is $a \gg f_i$. Furthermore, we
can set $\eta=1$ by choosing a perturbation with $\delta u =
\delta v$. Indeed, we only want to reproduce the divergence
$1/({\rm log}\,|\Delta|)^3$ associated to the `physical'
discriminant component $\Delta=0$. Expanding for $a$ large, the
expression for $S_i$ in terms of $\Delta$ is
\begin{equation} \label{eq:Sad}
S_1 \approx S_2=\frac{i \pi}{4}- i \pi \big(\frac{27}{4}+\Delta \big)\,.
\end{equation}
Also, $S_1-S_2\sim \mathcal O (a^{-3/2})$. Therefore, to leading
order in $a$, $S_1=S_2$ and they depend linearly on $\Delta$; up
to a shift by a constant, the integral $0 \le |\Delta| \le
\Lambda_f^3$ is hence translated to $0 \le |S_i| \le \Lambda_f^3$.

In this case, the gauge vacua formula (\ref{eq:gaugeNfinal}) reads
$$
N_{vac}^O(L_*)= \sum_{L=0}^{L_*}\;\sum_{(N_1,\,
N_2):\,gcd(N_i)|L}\,N_1^2 \,N_2^2
\;\sum_{(n_R,\,n_{NS})\,coprime}\;\times
$$
\begin{equation}\label{eq:NvacAD}
\times \,(n_R\,n_{NS})^2\,\Theta \Big(\frac{2\pi L}{{\rm
log}\,(\Lambda_0/\Lambda_f)^3}-(n_R^2+n_{NS}^2)(N_1^2+N_2^2) \Big)\,.
\end{equation}
A numerical evaluation shows that (\ref{eq:NvacAD}) has the same
dependence as (\ref{eq:IAD2}):
\begin{equation}\label{eq:NvacAD2}
N_{vac}^O(L_*) \approx \frac{2 \pi^2}{({\rm
log}\,(\Lambda_0/\Lambda_f)^3)^2}\;0.0235\,L_*^{4.060}\,,
\end{equation}
for $L_* \approx 1000$. Subleading corrections should be taken
into account, but their general dependence is hard to estimate.

This is a nontrivial check for the argument that we can map any
complicated singularity to the conifold regime and equivalently
count vacua there. Moreover, $n=2$ is the smallest genus for which
the symplectic transformations $Sp(2n-2,\mathbb Z)$ come into play
to count gauge vacua.

%%%%%%%%%%%%%%%%%%%%%%%%%%%%%%%%%%%%%%%%%%%%%%%%%%%%%%%%%%%%%%%%%%%%%
%%%%%%%%%%%%%%%%%%%%%%%%%%%%%%%%%%%%%%%%%%%%%%%%%%%%%%%%%%%%%%%%%%%%%
%%%%%%%%%%%%%%%%%%%%%%%%%%%%%%%%%%%%%%%%%%%%%%%%%%%%%%%%%%%%%%%%%%%%%
%%%%%%%%%%%%%%%%%%%%%%%%%%%%%%%%%%%%%%%%%%%%%%%%%%%%%%%%%%%%%%%%%%%%%
\section{Conclusions} \label{sec:concl}

In this paper we have shown that the number of supersymmetric
vacua $N_{vac}$ around ADE singularities of Calabi-Yau's in type
IIB flux compactifications is finite. The argument is based on the
existence of dual gauge theories, where finiteness may be shown.

Such singularities can be embedded in the noncompact CY
(\ref{eq:cy1}) and it is crucial that some of the fields become
nondynamical (couplings). The moduli are stabilized by turning on
both RR and NS flux through the compact cycles. We then perform a
geometric transition to connect this to the open string side.

The gauge theory is based on 5-brane bound states wrapping the
resolved 2-cycles. Its main properties are obtained by applying
S-duality to the DBI action on the resolved background. In
particular, the theory has fractional gauge couplings $\tau_i$;
the couplings are independent and hence cannot come from a UV
theory which is the usual $N=1$ SYM with superpotential $W(\Phi)$.

More importantly, the effective superpotential of the field theory
depends holomorphically on the couplings $(a_k)$. The dimension of
the chiral ring ($N_{vac}$) is thus invariant under changes
$\delta a_k$. We used this property to map a generic point in
field space $(S_i)$ to the conifold limit, while preserving the
number of vacua. In this semiclassical limit we showed that
$N_{vac}$ is finite.

Finally, we computed explicitly this number for the two simplest
singularities, namely the conifold point and the $n=2$
Argyres-Douglas point. The results from the gravity side and gauge
side match. This agreement is nontrivial since it involves quite
different concepts on both sides.

Let us compare both formulas. The gravity formula $\int {\rm
det}(-R-\omega)$ relates supersymmetric vacua to the geometry of
the moduli space. A simple topological interpretation~\cite{ashok}
is that it gives the number of zeroes of the section $D_i W_{eff}
\in \,\Gamma(\mathcal M \Omega \otimes \mathcal L)$. Clearly, it
is explicitly invariant under symplectic transformations. However,
the analysis of singularities is not straightforward, in
particular because the blow-up procedure becomes very involved as
we analyze higher codimension singularities.

On the other hand, the physics underlying the gauge theory result
is that of fractional instantons, bound states of 5-branes, RG
flow of the gauge couplings and combinatorics between the matrix
model cuts. The formula is explicitly finite after the mapping to
the semiclassical region. As a result, we recognize the exponent
$p$ in (\ref{eq:eguchi}) as the degree of the tree-level gauge
superpotential $W(\Phi)$ in which the singularity may be minimally
embedded. We should nevertheless point out that for $n \ge 3$
there remain symplectic generators that have to be fixed by
further restricting the fluxes to a fundamental region and this is
in general complicated. Another issue is that the combinatorics
grows very rapidly with $n$ and numerical computations become more
difficult.

A technical point that could be addressed in the future is to
understand better the origin of subleading corrections to the
gravity formula. These appear because the flux space is in fact a
lattice. The gauge theory approach might help in this direction.

We should note that the present results are based on the duality
between the closed (deformed) and open (resolved) sides. We
haven't been able to fully prove this, although we did show that
both sectors have the same IR physics. It would be very
interesting to continue this, perhaps with a supergravity analysis
along the lines of \cite{dasg}. If a lift to M-theory is possible,
the geometric transition might reduce to a duality between M5
branes, as in the Dijkgraaf-Vafa context.

\vskip 4mm
\subsection*{Acknowledgments} First I would like to thank my
advisor M. R. Douglas for suggesting this problem and for his
support and guidance throughout all the stages of the project.
From the beginning I have greatly benefited from extensive
explanations and discussions with F. Denef, D.-E. Diaconescu and
B. Florea. It is also a pleasure to thank G. Aldazabal, D. Belov,
C. D. Fosco, S. Franco, J. Juknevich, S. Klevtsov, S. Lukic, A.
Nacif, S. Ramanujam, K. van den Broek and A. Uranga for many
interesting discussions and suggestions. This research is
supported by Rutgers Department of Physics.

\end{document}